# Locomotion of Active Sheets Driven by Curvature Modulation

**Authors:** Omri Y. Cohen[1], Yael Klein[1], Eran Sharon[1]*

**Affiliations:**

[1]The Racah Institute of Physics, The Hebrew University of Jerusalem; Jerusalem, 91904, Israel.

*Corresponding author. Email: erans@mail.huji.ac.il

**Abstract:** The locomotion of flexible membrane-like organisms on top of curved surfaces appears in different contexts and scales. Still, such dynamics have not yet been quantitatively modeled and no realization of such motion in manmade systems has been achieved. We present an experimental and theoretical study of active gel ribbons surfing on a curved fluid-fluid interface via periodic modulation of their reference curvature. We derive a theoretical model, in which forces and torques emerge from curvature mismatch between the ribbon and the substrate. Analytic and numerical solutions of the equations of motion successfully predict the experimentally measured velocity profiles. We conclude by highlighting the relevance of this new, curvature-driven, mode of locomotion for a broad range of mechanical, as well as biological systems.

**Main Text:** Locomotion is a challenging task. The relevant object has to metabolize energy, to convert it to shape changes that interact with the surrounding environment and generate (often periodically) forces that lead to the accumulation of momentum. Locomotion seems especially challenging for membrane-like structures, such as cells and slender tissues that are attached to a surface. How can they generate net forces and torques for locomotion? Can such forces be generated without application of locally tangential stresses and if so, how these are related to the curvature of the surface and to the membrane properties? Experimental observations show that cells move and orient themselves in correlation with the substrate's curvature (*1-4*), a phenomenon that is known as curvotaxis. Theoretical modeling of such dynamics analyzes stress distribution that is generated by the cell and interacts with the surface (*5,6*). Still, the understanding of curvotaxis is qualitative and no equation of motion for cell propagation were derived.

Curvature driven locomotion occurs also on a much larger scale and different, seemingly unrelated, physical context. For example, on scales of mm-cm, slender organisms can propagate on curved fluid interface by varying their curvature (*7*). It is well known that objects that float on top of a liquid interface experience forces in the vicinity of a boundary, where the curved fluid meniscus breaks the translation symmetry. Such forces are commonly explained to result from asymmetric capillary stresses that lead to a net force on the floating body. These forces can be attractive or repulsive, depending on the body's shape, as was shown for solid floating objects (*8-10*). The dynamics becomes richer, when considering slender flexible floating objects (*11*).

We have constructed a synthetic curvotaxic system and studied its dynamics. It consists of active gel ribbons that harvest energy from the surrounding fluid and convert it into periodic shape changes, i.e., bending, while being confined to a curved interface between two liquids. The ribbons "surf" along the curved interface and we accurately measure their velocity profiles. Using the theory of incompatible elastic sheets, we constructed a theoretical model that describes this dynamic. The energy of the system stems from a mismatch between the curvature of the surface (the substrate) and the time dependent curvature of the ribbon. It typically consists of terms that result from deformations of the ribbon and from terms that are associated with deformation of the substrate. Gradients in the energy lead to forces and torques on the ribbon. We integrate the equations of motion and compute the trajectories of the ribbons. The theoretical model successfully recovers experimental results and allows predicting more general and more complex dynamics.

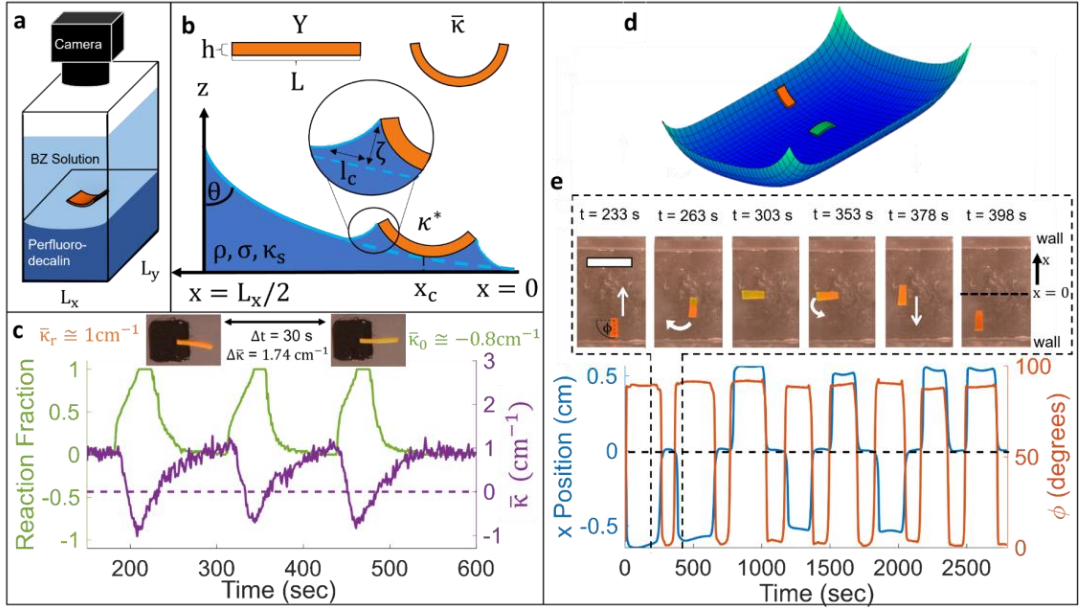

**Figure 1: The experimental setup and basic observations. a)** A BZ gel ribbon (L = 0.43 cm, h = 550 μm, w = 0.15 cm) is placed on the interface between two liquids: Perfluorodecalin ($\rho = 1.91$ g/cm$^3$) and BZ solution ($\rho = 1.03$ g/cm$^3$) in a rectangular glass aquarium ($L_x = 2.05$ cm, $L_y = 3$ cm). Its motion is tracked from above by a camera. **b)** An illustration of an elastic ribbon of length L, thickness h, Young's modulus Y and reference curvature $\bar{\kappa}$ floating on a fluid interface with curvature $\kappa_s(x)$. The density of the liquid is $\rho$ and its surface tension is $\sigma$. $\theta$ is the contact angle of the liquid with the wall and $l_c$ is its capillary length. Once the ribbon is placed on the surface, it pulls up fluid to a maximal height $\zeta$. The location of the ribbon's center is denoted by $x_c$ and the origin is at the center of the container. **c)** The reaction fraction (green line) and the curvature of a BZ gel ribbon (purple line) as functions of time. The ribbon is fully immersed in BZ solution (not confined to an interface). Periodic behavior is observed. The reference curvature oscillates between $\bar{\kappa}_r = 1$ cm$^{-1}$ and $\bar{\kappa}_o = -0.8$ cm$^{-1}$ in correlation with the reaction phase. **d)** An illustration of the two stationary states of a gel ribbon on top of the curved interface. The curvature of the surface is that of the 2D meniscus, which is positive everywhere. When the ribbon's curvature is positive (orange state), it stays near the wall, where its curvature best fits the positive surface curvature, pointing towards the wall. When its curvature is negative (green state), there is no position in which ribbon's and surface curvature match. To minimize the energy, the gel selects the flattest position and orientation on the surface – the center of the container, with orientation along the long direction of the container. **e)** The position (x, blue line) and orientation (ϕ, orange line) of the ribbon as functions of time. The ribbon moves periodically between the center of the container and its walls, while simultaneously changing its orientation. Dashed frame: images of one period of motion. The angle ϕ, as well as the two container's walls are indicated on the left and right panels, respectively. scale bar: 1 cm.

Our experimental system (Figure 1a) consists of a rectangular glass container, filled with two stratified liquids: a few centimeters of Perfluorodecalin ($\rho = 1.91$ g/cm$^3$) covered with a double amount of BZ solution ($\rho = 1.03$ g/cm$^3$). A narrow strip of BZ gel (see Supplementary Materials for detailed explanation on BZ gels and BZ solution) is placed on the interface between the two liquids. While the strip is immersed in the BZ solution, periodic reaction fronts propagate inside the gel and cause it to change its color from orange to green. As shown in (*12,13*) the equilibrium volume of the gel is coupled to the phase of the reaction: In the green (oxidized) phase, the gel swells in a factor 1.4 compared to its volume in the orange (reduced) state. This determines a reference geometry, which periodically oscillates in time and space. Our ribbons are asymmetric across their thickness (see SM) with enhanced reaction in one side compared to the other. As a result, the dominant effect of the reaction consists of periodic changes in the ribbons' reference curvature $\bar{\kappa}$, which oscillates between positive and negative values (Figure 1c and Movie S1). When the BZ gel strips are left to float freely on the fluid interface, they perform a periodic motion between the walls and the center of the container (Figure 1d and Movies S2-S3) We follow the motion (see details in SM) by a camera that is placed above the container. The motion consists of the following stages: The gel is initially in its orange reduced phase, "parking" next to one of the container's walls, pointing towards it. When the BZ reaction starts, the color of the gel begins to change to green, and it starts to move towards the center of the container. As the reaction wave continues to spread, close to the center of the container, the gel rotates. Eventually, as the gel is fully oxidized, it stops at the center of the container oriented perpendicularly to its original orientation (Fig. 1d). When the reaction decays and the gel turns orange again, it rotates and surfs towards the closest wall where it stops. This completes a full period of the motion, until the next reaction cycle begins[1]. This periodic motion can persist for hundreds of periods, as long as the BZ solution is kept fresh. We aim to explain the origin of this motion, its characteristics and to discuss to what extent it is general.

We model our system within the framework of elastocapillarity: The coupling between capillary forces and the elasticity of thin sheets was shown to generate different novel phenomena, such as wrapping (*14,15*) and bundling (*16*). We will show, however, that our modeling is relevant to a much wider class of phenomena. We start by studying the problem, which is described in Fig.1. It consists of a flexible ribbon, which is confined to the interface between two liquids with density difference, $\Delta\rho$ and interfacial surface tension $\sigma$. The ribbon has a *fixed* uniaxial reference curvature, $\bar{\kappa}$, along its long dimension, length L, width $w < L$, thickness $h \ll w$, and Young's modulus Y. The fluid interface is not flat. Near the container's wall, it develops a meniscus, with a characteristic height function $z(x,y)$ and reference curvature tensor $\mathbf{b_s}$. Far enough from the container's corners, $z(x,y) = z(x) \approx l_c \cot(\theta) e^{-(x-L_x/2)/l_c}$ and the reference curvature is uniaxial $\kappa_s(x)$. Here, the container's walls are located at $x = \pm L_x/2$, i.e., $x = 0$ is at the center of the container, $\theta$ is the contact angle and $l_c = (\sigma/\Delta\rho g)^{1/2}$, the capillary length. When the ribbon is placed on top of a two fluids interface, both the strip and the interface are deformed. The ribbon pulls up some liquid to a maximal height $\zeta$ (see Figure 1b) - a deformation that coasts gravitational energy. On the other hand, the liquid pulls back and bends the ribbon off its reference curvature - a deformation that costs elastic bending energy. The actual curvature of the ribbon, as well as its motion, are dictated by the balance between these two competing energies. Using the theory of non-Euclidean sheets (*17-19*), we can write the elastic energy functional of the ribbon which consists of a stretching term $E_S$ and a bending term $E_B$:

$$E_{el}[\mathbf{a}, \mathbf{b}] = Y \int \left[ h E_S (\mathbf{a} - \bar{\mathbf{a}}) + h^3 E_B (\mathbf{b} - \bar{\mathbf{b}}) \right] dA \quad (1)$$

Where $\mathbf{a}$ is the metric tensor and $\mathbf{b}$ is the curvature tensor of the ribbon. $\bar{\mathbf{a}}$ and $\bar{\mathbf{b}}$ are the reference metric and curvature tensors, respectively. For our system, in which the reference curvatures of both the ribbon and the surface (far from the container's corners), $\bar{\kappa}$ and $\kappa_s$ respectively, are uniaxial, this expression becomes (see SM):

$$E_{el} = Y h^3 L (\kappa(x, \phi) - \bar{\kappa})^2 \quad (2)$$

Where $\phi$ is a rotational degree of freedom, with $\phi = \pi/2$ is along $\hat{x}$ (see Figure 1e) and $\kappa(x, \phi)$, the actual ribbon's curvature at $(x, \phi)$. The energy associated with the deformation of the substrate to a profile $\xi(x)$ that describes the height of the interface above the unperturbed interface, $z(x)$, is approximately:

$$E_{sub} = \Delta \rho g \int \left( \xi(x') - z(x') \right)^2 dx' \approx \Delta \rho g l_c \zeta^2 \quad (3)$$

For small deformations $\zeta \approx L^2 (\kappa - \kappa_s)$, which yields for the 2D case (see SM):

$$E_{sub} \cong \Delta \rho g l_c L^4 \text{Tr}[\mathbf{b}(x, y, \phi) - \mathbf{b_s}(x, y)]^2 \quad (4)$$

---
[1] In a similar experiment that was done with a square container, the gel moved from the center to three out of the four walls and had no preferred orientation in the vicinity of the center (Movie S4).

The total potential energy of the ribbon is $E = E_{el} + E_{sub}$ and the selected curvature $\kappa^*$ at every point is determined by minimization of this energy with respect to $\kappa$. Substituting $\kappa^*(x, \phi)$ back into the energy expression yields a *space dependent potential energy*.

In the one-dimensional case, i.e., when **b** has only one non-zero entry, there is an analytic expression for $\kappa^*(x)$:

$$\kappa^*(x) = \frac{\bar{O}\frac{\kappa_s(x)}{\bar{\kappa}} + 1}{(\bar{O} + 1)}\bar{\kappa} \tag{5}$$

Where $\bar{O} = \Delta\rho g l_c/Y \cdot (L/h)^3 = (\Delta\rho g \sigma/Y^2)^{1/2} \cdot (L/h)^3$ is a dimensionless number that represents the ratio between the stiffness of the substrate (the liquid interface) and that of the ribbon. The $\bar{O} \ll 1$ limit corresponds to stiff ribbons. In this limit, the ribbon's curvature is approximately constant and equals to $\bar{\kappa}$. In this case, most of the system's energy is the substrate energy and the system behaves like rigid floating bodies that have been studied before (*9,20*). On the other hand, if the ribbon is very flexible, namely $\bar{O} \gg 1$, it is fully confined to the surface, and its curvature becomes $\kappa_s$. In this limit the substrate energy is negligible. All the dynamics is dominated by the elastic energy of the ribbon. The mechanism of the motion is now clear: a curved elastic ribbon confined to a non-uniformly curved surface moves towards the point where its curvature fits the curvature of the surface. By taking the spatial and angular derivatives of the energy, one can obtain the forces and torques that act on the ribbon. Using the fact that the system is overdamped, we can even go further, and derive the equation of motion of the ribbon: In this case, the linear and angular velocities, $v$ and $\omega$, are proportional to the forces ($\vec{F}$) and torques ($\vec{M}$) that act on the ribbon respectively:

$$\vec{v}(x, y, \phi) = \vec{F}(x, y, \phi)/D = -\vec{\nabla}E(x, y, \phi)\big|_{\kappa=\kappa^*}/D$$

$$\vec{\omega}(x, y, \phi) = \vec{M}(x, y, \phi)/I = -\frac{\partial E}{\partial \phi}\hat{z}\big|_{\kappa=\kappa^*}/I \tag{6}$$

, The linear and angular drag coefficients, D and I, depend on the ribbon's dimensions and fluid viscosity, and contain numerical factors that are missing in our expressions for both energy terms.

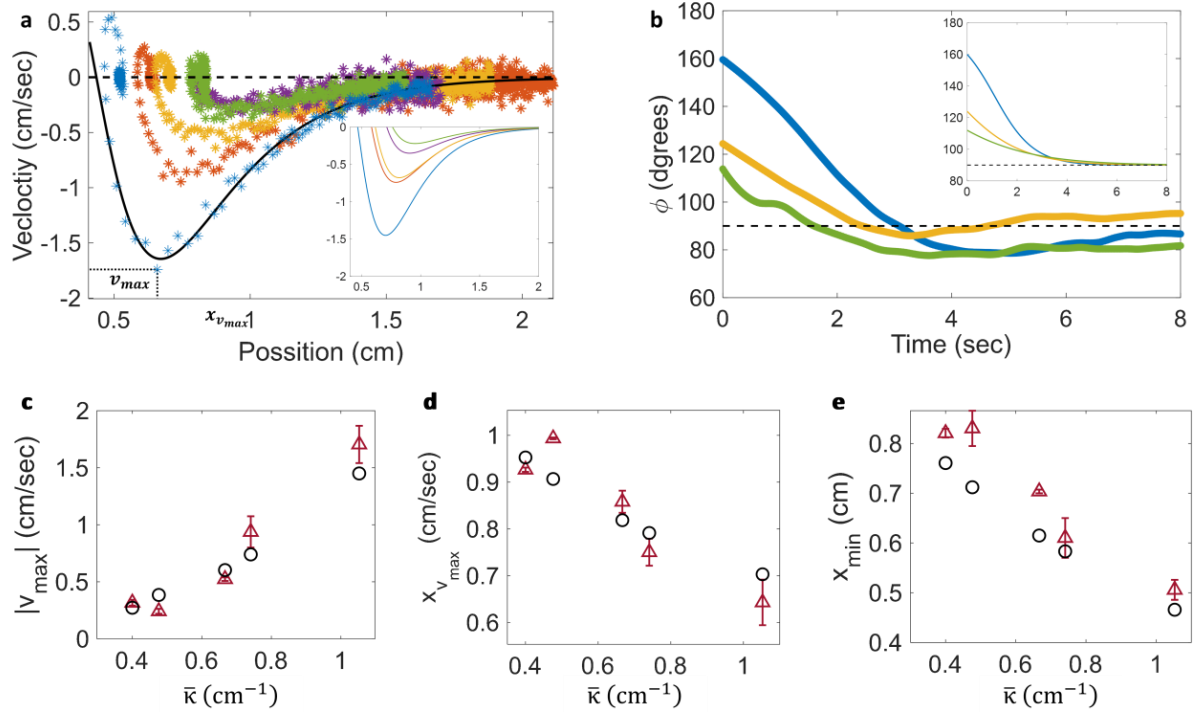

**Figure 2: The effect of reference curvature on the motion of an elastic ribbon.** Velocity profiles (**a**) and orientation (**b**) of 5 different ribbons (w = 0.3 cm, L = 0.7 cm) floating on water and attracted to a wall which is located at x = 0. The reference curvatures and thicknesses are $1.05$ cm$^{-1}$ and 290 μm (blue), $0.74$ cm$^{-1}$ and 290 μm (orange), $0.67$ cm$^{-1}$ and 290 μm (yellow), $0.48$ cm$^{-1}$ and 380 μm (purple) and $0.4$ cm$^{-1}$ and 380 μm (green). The solid black line in (**a**) is a fit according to Eq. 6 to the measured trajectory of the ribbon with the largest curvature (blue symbols), with D = 49 (dyn · sec)/cm$^2$ and θ = 31°. Insets: Theoretical velocity and orientation curves of the ribbons (Eq. 6) with the relevant experimental parameters and the average values of the contact angle and drag coefficients: θ = 28°, D = 55 (dyn · sec)/cm$^2$ and I = 0.75 dyn · sec. The qualitative agreement between theory and experiments is clear. The absolute value of the maximal velocity (**c**), the location of the maximal velocity (**d**) and the stopping points (**e**) for the data in panel (**a**), as measured experimentally (red triangles) and computed using Eq. 6 (black circles).

In order to verify our model predictions, we conducted a series of experiments with uniformly curved ribbons made of Polyvinylsiloxane (PVS) (see SM). The ribbons were placed on the surface of distilled water approximately 2 cm from the straight container wall. Their motion was tracked by a camera that was placed above the container. From the pictures, we extracted x(t), the location of the center of the ribbon and ϕ(t), its orientation, over time. The velocity curves (Fig. 2a) are characterized by an acceleration towards the wall, a maximal velocity, $v_{max}$, which is obtained at some position, $x_{v_{max}}$, and then deceleration until the ribbon stops at a distance $x_{min}$, from the wall. Our model (solid line) is successful in describing the entire motion for $\bar{\kappa} = 1.05$ cm$^{-1}$ and h = 290 μm. The inset shows additional numerical solutions of the equation of motion that correspond (same colors) to the experimental parameters and fixed values of θ = 28° and D = 55 (dyn · sec)/cm$^2$ which are the average values of all best fitting parameters (the std in θ and D are 4° and 16 (dyn · sec)/cm$^2$ respectively. See SM for more details). Both experimental and theoretical curves show that as $\bar{\kappa}$ increases, the maximal velocity increases, it is obtained closer to the wall and the gel stopping point is also closer to the wall. Following the evolution of the ribbon orientation (Fig. 2b) we see that the initial orientation, which was chosen to be nearly parallel to the wall ($\phi = 180°$ corresponds to a parallel orientation), decreases, as the ribbon spins, orienting itself perpendicularly to the wall ($\phi = 90°$) *before* starting its linear motion towards the wall (Movie S5). The success of the modeling can also be seen in the supplemental movies that present theoretical trajectories (Movie S6). Note that in all cases, the ribbons oscillate a few times before settling at their equilibrium points. This indicates that the assumption of negligible inertia term is not valid at the final stage of the motion.

The theoretical model provides a prediction for the variation of the velocity profiles with different parameters of the ribbon. The predictions for the variation due to changes in the reference curvature (which is the change that occurs in the BZ gels) are verified by comparing the measured and the predicted values of $|v_{max}|$ (Fig. 2c), $x_{v_{max}}$ (Fig. 2d), and $x_{min}$ (Fig. 2e), for different values of $\bar{\kappa}$. The model provides a good description of the trajectories, using fixed values of all parameters. In addition, it provides good predictions for the dependence of the trajectories on L and Y (see SM).

We now turn to analyze the motion of the active BZ gel strips. Here, the reference curvature varies in time, $\bar{\kappa} = \bar{\kappa}(t)$, with a typical period of $\sim 150$ s (Fig. 1c, Movie S1) and a simple integration of the equation of motion is not possible. We superimpose a sequence of full trajectories (Fig. 3a), in which the strip approaches the walls and retracts back to the center of the container. We note that the attraction trajectories i.e., motion from the center, $x = 0$, towards the boundaries, $x = \pm 1.025$ cm, and the opposite, repulsion trajectories, are different: In our experiments, the attraction maximal velocity was 2-3 times higher than the maximal repulsion velocity. In addition, the shape of the velocity profile is qualitatively different: the repulsion profile is more symmetric (with respect to x) than the attraction profile. As for the angular velocity, the ribbon rotates during the repulsion stage, when its maximal angular velocity is gained in the vicinity of the container center (for example $x = 0.6$ mm, in Figure 3.b). On the other hand, in the attraction phase, the ribbon completes its rotation and only then starts to move back towards the wall, while its angular velocity is zero. To understand these differences, we note that in the attraction phase, the uniaxial curvature along the ribbon generates attractive forces, only in orientations close to $\phi = 90^0$. Therefore, rotation precedes linear acceleration. During the repulsion, the ribbon is already oriented along $\phi = 90^0$. Therefore, small changes in curvature generate linear acceleration and the ribbon reaches the central parts of the container before $\bar{\kappa}(t)$ attains its minimal value. The asymmetries in the linear and angular velocity curves demonstrate the complexity of the relevant dynamics, which results from the existence of two independent time scales, the intrinsic timescale, τ and the dynamical timescale - the typical traveling time of the strip, and two degrees of freedom ($\phi, x$).

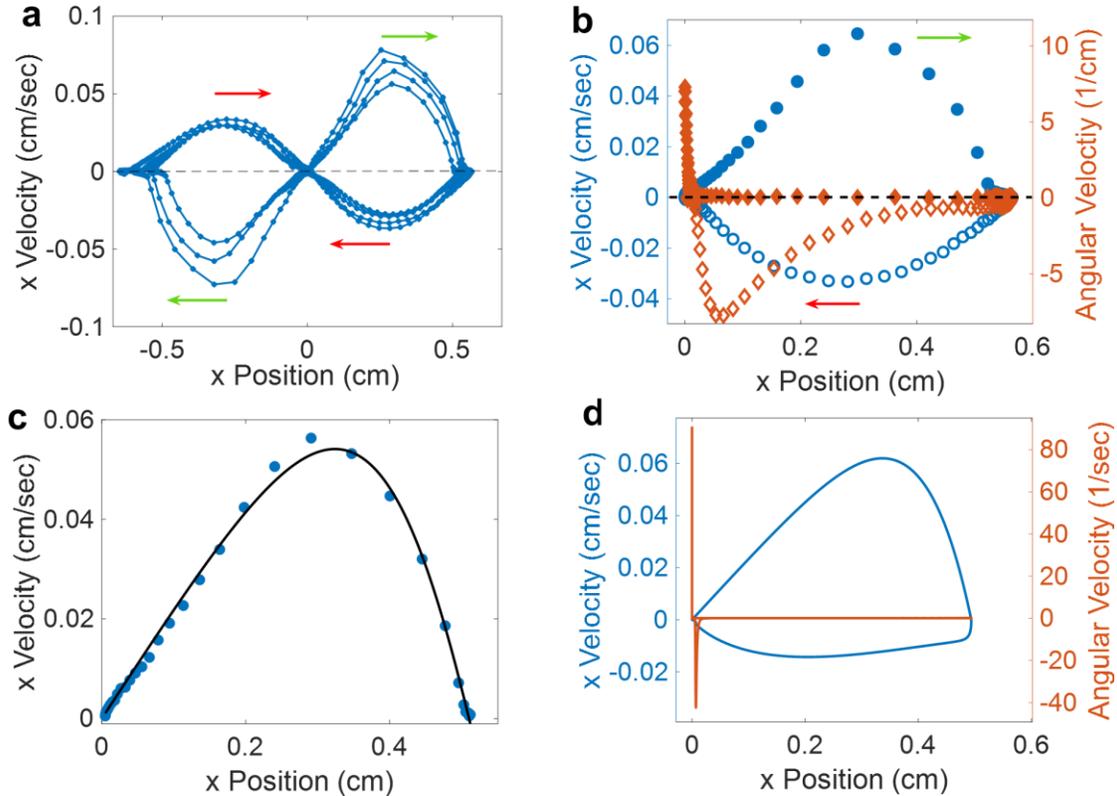

**Figure 3: Velocity profiles of active BZ gel ribbons. a)** Velocity of a BZ gel ribbon (L = 0.43 cm, h = 550 μm, w = 0.15 cm) as function of its position, during seven complete cycles of the motion. The origin is in the center of the container and the walls are at $x = \pm 1.025$ cm. Green and red arrows refer to motion from the center to the walls and from the walls to the center, respectively. **b)** The linear (blue circles) and angular (orange diamonds) velocity profiles as measured over one cycle. The full and empty symbols correspond to the attraction and repulsion profile, respectively. **c)** A typical measured attraction profile (blue stars) with a theoretical curve (black line) (Eq. 6) fitted to it, with $\theta = 25°$, $l_c = 0.25$ cm and D = 455 (dyn · sec)/cm². **d)** Theoretical linear and angular velocity profiles, using the parameters obtained in Fig. 1 and Fig. 4c: $\bar{\kappa}_r = 1$ cm$^{-1}$, $\bar{\kappa}_o = -0.8$ cm$^{-1}$, $\dot{\bar{\kappa}} = 0.06$ cm$^1$/sec, D = 418 (dyn · sec)/cm², $l_c = 0.26$ cm, $\theta = 27°$, the measured L, h, Y, and I = 5 dyn · sec as a free parameter.

Since the acceleration toward the wall takes place while $\bar{\kappa}$ is approximately constant (the gel is in its reduced phase), we can analyze the attraction trajectory, using Eq. 6. The fit was done using $\bar{\kappa} = \bar{\kappa}_r = 1$ cm$^{-1}$ (see Fig. 1), while D, $\theta$ and $l_c$ were free parameters. The theoretical curve fits the measured trajectory well (Fig. 3c). We use seven attractive trajectories to extract the average values of D, $\theta$ and $l_c$. These parameters, together with $\bar{\kappa}(t)$ are used for numerical derivation of the repulsion and angular velocity curves (Fig. 3d). As in the experiments (Fig. 3b), the attraction and repulsion profiles, of both linear and angular velocities, are quantitatively different. The simulation captures the fact that during the attraction stage, the ribbon first spins and only then starts its acceleration towards the wall, and the rotation occurs at the center of the container. The movie that presents the theoretically computed motion (Movie S7) indeed resembles the experimentally measured motion (Movie S2). As for the angular velocity, the computation captures only part of the profile's characteristics. Though there is a difference between the attractive and repulsive trajectories, the repulsive profile is more located compared to the measured one, and the maximal velocities are higher.

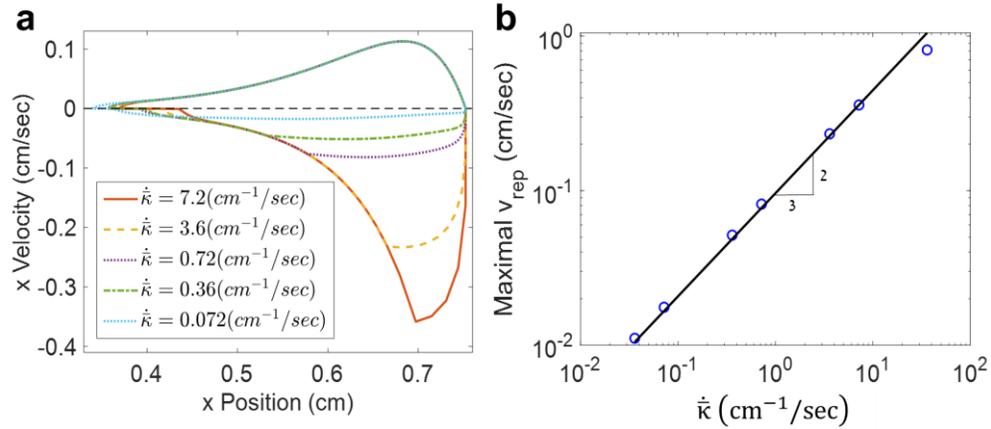

**Figure 4: The effect of the reaction rate on the repulsion profile. a)** Numerical velocity profiles for different curvature change rates, $\dot{\bar{\kappa}}$. **b)** A log-log plot of the maximal repulsion velocity as function of $\dot{\bar{\kappa}}$. The numerical data is consistent with the dimensional analysis predictions: $v_{max} \sim \dot{\bar{\kappa}}^{2/3}$ (solid line). The best power law fit to the data is $\dot{\bar{\kappa}}^{0.63}$.

As a last stage of our analysis, we study the interplay between the intrinsic time scale, $\tau_{\bar{\kappa}}$, the time of the gel oxidation process (see Fig. 1c), and the time of the motion itself $\tau_v$. We solved numerically the equations of motion, while changing $\tau_{\bar{\kappa}}$ over three orders of magnitude (0.05 – 50 sec). The velocity profiles are shown in Figure 4a. As expected, the attraction profiles ($v_x > 0$) in all cases are the same, since in all of them, the propagation occurs when the curvature is already fixed to its minimal value, $\bar{\kappa}_r = -1 \text{ cm}^{-1}$. In contrast, the repulsion profiles ($v_x < 0$) vary a lot. Each of them is composed of two parts: as long as the reference curvature keeps changing the velocity increases concavely. At this stage, the velocity reaches its maximal value. Once $\bar{\kappa}$ becomes constant, the velocity decays convexly in a profile that is common to all samples and is described by Eq. 6. For small $\dot{\bar{\kappa}}$, the concave part lasts for most of the repulsion, and we get symmetric profiles that resemble the profiles we measured experimentally (Figure 3b). Indeed, for $\dot{\bar{\kappa}} = 0.072 \text{ cm}^{-1}/\text{sec}$ (the cyan curve), which most closely corresponds to the experimental parameters, the velocity profile is similar to the experimental curves. As $\dot{\bar{\kappa}}$ increases, the velocity profile becomes asymmetric and the maximal velocity increases. To further investigate the effect of the two timescales on the velocity profile, we plot the maximal repulsion velocity as function of $\dot{\bar{\kappa}}$ (Figure 4b). We find a power law dependence over more than three decades in $\dot{\bar{\kappa}}$. This functional dependence can be explained with a simple dimensional analysis, which states that for every profile, the relation $\tau_v \approx \tau_{\bar{\kappa}}$ must hold (i.e., the travel from the wall to a distance of order $l_c$, is completed within a time of order $\tau_{\bar{\kappa}}$). This leads (see SM) to the relation $v_{rep} \sim \left(\frac{\rho g L^4 l_c^2}{D}\right)^{1/3} \dot{\bar{\kappa}}^{2/3}$, which is consistent with our data.

**Discussion**

In this work we presented and studied the locomotion, driven by curvature modulation (curvotaxis) of synthetic active strips. We synthesized ribbons made of active BZ gel, designed to periodically alternate their reference curvature, and measured their propagation along a curved fluid interface (meniscus). We developed an energetic approach to analyze and model this motion. The motion is driven by gradients in the energy that stem from geometrical incompatibility between the surface and the ribbon. The validity of the model was verified by comparing experiments in passive as well as active sheets to analytical and numerical results. The motion is governed by two time scales: The intrinsic time scale, $\tau_{\bar{\kappa}}$, which is the period of curvature oscillations, and the dynamic time scale, $\tau_v$, the ratio between typical length and velocity scales of the motion. The ratio between these time scales qualitatively changes the characteristics of the velocity profiles. In the current work the motion stems only from uniaxial bending of the ribbon and the surface was a fluid. We suggest, however, that our work has a far-reaching impact, and, in a sense, it is the tip of the iceberg. It can be extended in both, the type of incompatibility and the type of substrate material.

*Types of incompatibilities:* In our experiments, the mismatch in the uniaxial curvature introduces the dimensionless number, $\bar{O}$, that characterizes the ratio between the surface stiffness (determined by the gravitational energy of the lifted liquid) and the ribbon's bending stiffness. For a two-dimensional sheet, with a nonnegligible width w, the reference curvature is typically a two-dimensional tensor $\bar{\mathbf{b}}$. Furthermore, in this case, the stretching energy term (Eq. 1) could play a role, as a conflict between $\mathbf{a}$ and $\bar{\mathbf{a}}$ is possible. Hence, the total energy of an elastic sheet that floats on a curved fluid interface will be composed of three terms: a substrate deformation term, which in our case scales like $\rho g l_c w L^4 (\mathbf{b} - \mathbf{b_s})^2$, a bending term that scales like $Y h^3 L w (\mathbf{b} - \bar{\mathbf{b}})^2$ and the new, stretching, term that scales like $Y h w L (\mathbf{a} - \bar{\mathbf{a}})^2$. In this case we expect an additional two dimensionless numbers that characterize the dynamics. For a square sheet (w = L) with a reference metric that determines a reference Gaussian curvature $\bar{K}$, we find (see SM for details): $\widetilde{O} = \frac{\Delta \rho g l_c}{Y} \frac{\bar{\kappa}^{-2}}{Lh} = \sqrt{\frac{\Delta \rho g \sigma}{Y} \frac{\bar{K}^{-1}}{Lh}}$ which describes the ratio between the stretching stiffness and the substrate (fluid) stiffness, and $A = \frac{h^2}{L^4 \bar{\kappa}^2} = \frac{h^2}{L^4 \bar{K}} = \frac{\widetilde{O}}{\bar{O}}$ that characterizes the ratio between the bending and stretching stiffnesses, and is familiar from studies of free frustrated sheets (*18*). By substituting the values of h, L and κ, one can determine which of these terms dominates. For example, the case studied in (*21*) corresponds to the limit $A \ll \widetilde{O} \ll \bar{O}$. The two-dimensional phase space which rises from this analysis is much richer than the one we discuss in this work and requires a deeper investigation.

*Substrate material:* The energy term in Eq. 4 expresses the energy that results from deforming the substrate from its unperturbed configuration. In our system, the substrate is a fluid, and therefore, the deformation energy takes the form we used. However, flexible sheets could be confined to various types of deformable substrates, including soft solid substrates, as commonly occurs in biological systems. Following our approach, one could express the energy associated with deformation of every type of substrate. Different dimensionless numbers, equivalent to $\bar{O}$ and $\widetilde{O}$ will determine the dynamics in such systems. Regardless of the details of the problem, gradients in $\kappa_s$, will lead to net forces and torques on the sheet. A possible important example of such mode of locomotion is possibly found in microscopical systems, where cells and small tissue fragments (*4*) are observed to move (*1,3*) and orient themselves (*2,4*) in correlation with the substrate's curvature. We suggest that the formalism presented here provides new tools for quantitative study and analysis of such curvotaxis phenomena.

The locomotive principle presented in this work is in action even for infinitely rigid substrates. In such a case (which in our system corresponds to $\bar{O} \to \infty$), the substrate energy term is completely negligible and the energy of the system is composed only from the elastic, bending and stretching terms. Surprisingly, the elastic term alone converts gradients in $\kappa_s$ into forces and torques on the sheet, despite the fact that in such cases all forces are locally normal to the sheet.

Locomotion driven by curvature incompatibility is, therefore, a general phenomenon that can occur in different physical and biological systems. It has a broad range of parameters that can lead to various types of dynamics. Our work covers only a small part of interesting phenomena associated with this type of motion. Furthermore, as in our study, integration of the equation of motion provides the entire dynamics of such systems. Questions that are related to the efficiency of this mode of locomotion, as well as the possibility of chaotic dynamics are only two examples that should be studied in future experimental and theoretical works.

**Materials and Methods**

BZ gels ribbons and BZ solution

The BZ gels were fabricated according to Maeda et. al. (*22*). We first made two solutions: a solution of 280mg of N,N′-Methylenebis(acrylamide) (Bis-Acrylamide) dissolved in 10cc of Dimethyl 9 sulfoxide (DMSO), and a solution of 276mg AMPS dissolved in 20cc of water. We than took 780mg N-isopropylacrylamide (NIPA) and 81mg of ruthenium(II)tris-(2,2′-bipyridine) (Ru(bpy)$_3^{2+}$) and added 0.5cc of the BIS solution, 2.5cc methanol and 2cc of the AMPS solution. Finally, we added 1cc of a solution of 0.2M 2,2′-Azobis(2-methylpropionitrile) (AIBN) in toluene to the mixture.

To get thin sheets of gel, we injected the solution between two glass plates that were pressed tightly together. Before the injection, we put a silicon-rubber gasket (Smooth-Sil 940) between the plates, to control the thickness of the gel sheet, and covered one of the plates with a thin Teflon sheet. The plates were placed in an oven at 60°C for 20 hours for polymerization. The Ru(bpy)$_3$ monomer is hydrophobic, and hence, during the polymerization it easily migrates to the hydrophobic Teflon plate, while the AMPS migrates to the opposite hydrophilic glass. As a result, a nonuniform distribution of Ru(bpy)$_3$ along the thickness was formed. On the surface side where the content of hydrophilic AMPS is higher, the swelling ratio of the gel membrane becomes larger than that on the opposite side in the same gel.

After the polymerization, the gels were gently removed from the plates and were transferred into ethanol for five days. Afterword, they were placed in a series of ethanol-water baths of increasing water concentration (25%, 50%, 75%, 100%) for a full day each. The fully washed gels were cut with a razor to small (few mm long) rectangular ribbons.

Before each experiment, we prepared a "**BZ solution**" which consists of nitric acid (HNO$_3$) 0.88M, sodium bromate (NaBrO$_3$) 0.084M, and malonic acid (C$_3$H$_4$O$_4$) 0.062M. A few millimeters of Perfluorodecalin (C$_{10}$F$_{18}$) (density ρ = 1.91 g/cm$^3$)  were poured into a glass aquarium, and after that, a double amount of BZ solution was gently poured on top of it. The gel ribbons were transported to nitric acid for 5 minutes and then gently placed on the interface between the two liquids (Figure 1a). We tracked the dynamics with a Nikon camera that was placed above the aquarium. We took a photo every second.

Polyvinylsiloxane ribbons

Polyvinylsiloxane (PVS) is an elastomer that is fabricated by mixing a base polymer with a curing agent (i.e. a crosslinker) in equal measures. In our experiment, we used Elite Double 32, supplied by Zhermack (Italy). The Young's modulus of this elastomer is $Y \cong 700$ kPa , $7 \cdot 10^6$ dyns/cm² (*23*). We mixed equal amounts of base polymer and crosslinker and added to the mixture green water-based ink (5 drops for every 6 ml of mixture), to strengthen its green color. We poured the mixture on top of a series of glass cylinders in a range of radii (0.95 – 2.5 cm), let it flow under the effect of gravity until it covered the cylinder, and then left it to cure for more than 30 minutes. Due to the balance between the viscous stresses and gravity inside the mixture, the thickness of the obtained elastomer is relatively uniform (up to 8% for spheres, as was shown by Lee et al. (*24*)). Each cylindrical elastic sheet was cut by laser cutter into small ribbons of different lengths and widths. The thickness of each ribbon was measured by a caliper and its uniaxial curvature was calculated from the radius of the cylinder it was fabricated on.

The ribbons were placed gently on the water surface within a plastic container, a few centimeters from the container's boundary. The movement of these floating ribbons was tracked by a camera (Luminera model Lt225C, 40 frames per second) that was placed above the container.

**Supplementary Text**

Theoretical derivations

*The elastic energy of a thin floating ribbon*

As argued in the main text, the elastic energy functional can account for various scenarios of curvature incompatibilities, but our problem is simple; The reference Gaussian curvature of ribbon's metric is zero and so are those of the fluid surface (away from the container's corners) and the curvature, which is determined by $\bar{\mathbf{b}}$, is uniaxial. As a result, the stretching energy does not play a role and the only conflict is between $\bar{\mathbf{b}}$ and the surface curvature. Therefore, the elastic energy functional is:

$$E_{el}[\mathbf{b}] = Yh^3 \int \nu \text{Tr}^2[(\mathbf{b} - \bar{\mathbf{b}})] + (1 - \nu)\text{Tr}[(\mathbf{b} - \bar{\mathbf{b}})]^2 dA \quad (S.1)$$

Where $\nu$ is Poisson ratio. For constant curvature $\kappa$ along its long dimension and $\nu = 0$ the curvature tensors are:

$$\mathbf{b} = \kappa \begin{pmatrix} \sin^2(\phi) & \cos(\phi)\sin(\phi) \\ \cos(\phi)\sin(\phi) & \cos^2(\phi) \end{pmatrix}, \quad \bar{\mathbf{b}} = \bar{\kappa} \begin{pmatrix} \sin^2(\phi) & \cos(\phi)\sin(\phi) \\ \cos(\phi)\sin(\phi) & \cos^2(\phi) \end{pmatrix} \quad (S.2)$$

Where $\kappa$ and $\bar{\kappa}$ are the uniaxial actual and reference curvatures of the ribbon and $\phi$ is a rotational degree of freedom, with $\phi = \pi/2$ is along $\hat{x}$ (see Figure 1.d). Since the actual and reference curvatures orient together (or, in other words, the rotation matrix contributes nothing to the term $\text{Tr}[(\mathbf{b} - \bar{\mathbf{b}})]^2$),the final form of the elastic energy is therefore:

$$E_{el} = Yh^3 L(\kappa(x, \phi) - \bar{\kappa})^2 \quad (S.3)$$

*Approximating $\zeta$ using curvature difference*

In order to calculate $\zeta$, the maximal height of the lifted liquid, we used the following integral (in the one-dimensional case):

$$\zeta = \xi\left(x_c + \frac{L}{2}\right) - z(x_c) = \int_{x_c}^{x_c+\frac{L}{2}} \int_{x_c}^{u} (\xi''(u') - z''(u'))dudu' \quad (S.4)$$

where $x_c$ is the location of the center of the ribbon. We assume that the ribbon completely floats on the liquid and that its center is at the level of the liquid surface, meaning $\xi(x_c) = z(x_c)$ and $\xi'(x_c) = z'(x_c)$. Here, we also assume that the picture is symmetric around the center of the ribbon. This assumption is reasonable as long as the height difference between the liquid surface on both sides of the ribbons is small compared to its length. In this case, we can also approximate the curvatures of the surface, $\kappa_s(x)$, and the ribbon by the second derivative of their height functions:

$$\kappa_s(x) \cong z''(x), \qquad \kappa = \kappa(x) \cong \xi''(x) \quad (S.5)$$

Plugging this into the expression for $\zeta$ gives:

$$\zeta \cong \int_{x_c}^{x_c+\frac{L}{2}} \int_{x_c}^{u} (\kappa - \kappa_s(u'))dudu' \quad (S.6)$$

Again, under the assumption that the curvature of the surface changes slowly along the length of the ribbon, we can write:

$$\zeta \cong L^2(\kappa - \kappa(x_c))^2 \quad (S.7)$$

The generalization to the 2D case is immediate: we must replace the difference between the two scalar curvatures, to the difference between the curvature tensors $|\mathbf{b} - \mathbf{b_s}|$, where $\mathbf{b_s}(x, y)$ is the curvature tensor of the surface. Plugging it back to the expression for the substrate energy we get:

$$E_{sub} \cong \Delta\rho g l_c \zeta^2 \cong \Delta\rho g l_c L^4 |\mathbf{b}(x, y, \phi) - \mathbf{b_s}(x, y)|^2 = \Delta\rho g l_c L^4 \text{Tr}[\mathbf{b}(x, y, \phi) - \mathbf{b_s}(x, y)]^2 \quad (S.8)$$

*Modeling of a 2D compatible sheet*

In this case, we model a two dimensional sheet of length L and width w. Assuming that the two intrinsic and actual principal curvatures of the sheet at each point are of the same order, such that $\kappa_1 = \kappa_2 = \kappa$ and $\bar{\kappa}_1 = \bar{\kappa}_2 = \bar{\kappa}$, the actual and intrinsic Gaussian curvatures $K, \bar{K}$ can be written as $K = \kappa^2, \bar{K} = \bar{\kappa}^2$ and the total energy per unit area can be approximated as:

$$\frac{E}{wL} \sim \rho g l_c \left(L^3(\kappa - \kappa_{s,1})^2 + w^3(\kappa - \kappa_{s,2})^2\right) + 2Yh^3(\kappa - \bar{\kappa})^2 + Yhw^4(K - \bar{K})^2 \quad (S.9)$$

Where $\kappa_{s,1}$ and $\kappa_{s,2}$ are the principal curvatures of the surface. If the sheet is narrow and long, namely $w \ll L$, we recover our previous result. In the special case of a square sheet $L = w$, this expression becomes:

$$\frac{E}{L^2} \sim \rho g l_c L^3 \left((\kappa - \kappa_{s,1})^2 + (\kappa - \kappa_{s,2})^2\right) + 2Yh^3(\kappa - \bar{\kappa})^2 + YhL^4(K - \bar{K})^2 \quad (S.10)$$

*Dimensional analysis of the model*

We can identify three different timescales involved in the ribbon's motion. The first is simply the period of the curvature change $\tau_{\bar{\kappa}}$. Two other timescales are associated with the potential energy, one for each component of it. If $\bar{O} \gg 1$, the ribbon is fully confined to the surface and $\kappa \cong \kappa_s$. Hence, the elastic energy term is dominant, and the relevant timescale is:

$$\tau_{el} = \frac{l_c^2 D}{Yh^3 L \kappa_s^2} \quad (S.11)$$

On the other hand, if $\bar{O} \ll 1$ the curvature of the ribbon is similar to its intrinsic curvature $\kappa \cong \bar{\kappa}$, and the surface tension energy is more relevant. This time we can define:

$$\tau_{sub} = \frac{l_c D}{\rho g L^4 \bar{\kappa}^2} \tag{S.12}$$

In the numerical calculation related to Fig. 4, $\bar{O} = 0.0241$ and hence we can use $\tau_{sub}$. Using dimensional analysis, we find the scaling between the velocity and $\dot{\bar{\kappa}}$: The ribbon obtains its maximal velocity after it has traveled a length $l_{max}$. In our resualts, $l_{max}$ is of order of magnitude of $l_c$, and we can approximate v:

$$v = \frac{dx}{dt} \sim \frac{l_c}{\tau_{sub}} = \frac{\rho g L^4 \bar{\kappa}^2}{D} \tag{S.13}$$

On the other hand, the instantaneous $\bar{\kappa}$ of the ribbon can be approximated as $\bar{\kappa} \sim \dot{\bar{\kappa}} \tau_{\bar{\kappa}}$. In the calaulation, we saw that the $\tau_{\bar{\kappa}}$ is close to time in which the ribbon gained its maximal velocity. That allows us to set $\tau_{\bar{\kappa}} \cong \tau_{sub}$ and get:

$$\bar{\kappa} \sim \dot{\bar{\kappa}} \tau_{sub} \sim = \frac{l_c D}{\rho g L^4 \bar{\kappa}^2} \dot{\bar{\kappa}} \tag{S.14}$$

Extracting of $\bar{\kappa}$ from Eq. S.4 and substituting in Eq. S.3 yields:

$$v \sim \left(\frac{\rho g L^4 l_c^2}{D}\right)^{\frac{1}{3}} \dot{\bar{\kappa}}^{\frac{2}{3}} \tag{S.15}$$

Curvature modulation due to the BZ reaction and modeling

As the reaction propagates inside the gel, and it becomes greener, its curvature increases. When the reaction percentage is approximately 69%, the curvature becomes zero, and from that point on, it changes its sign and becomes positive. The variation of the curvature along time is not symmetric: its minimal value (when the gel is fully in the oxidized state) is $\bar{\kappa}_o \cong -0.8 \text{ cm}^{-1}$ but the maximal value (when it is fully in the reduced state) is $\bar{\kappa}_r \cong 1 \text{ cm}^{-1}$. The curvature remains at its minimal value for short time (~10 seconds) and then increases. Furthermore, the increasing time of the curvature is 2-3 times longer than the decreasing time (20-30 seconds compering to 40-60 seconds).

In order to solve numerically the equations of motion for a BZ strip, which has a curvature that changes over time, we had to model $\bar{\kappa}(t)$. Basing on the measurement of $\bar{\kappa}(t)$ (Figure 1.c), we set the curvature to change linearly from $\bar{\kappa}_r = 1 \text{ cm}^{-1}$ in the reduced state, to $\bar{\kappa}_o = -0.8 \text{ cm}^{-1}$ in the oxidized state and vice versa, such that the oxidation is two times faster than the reduction (Supplementary Figure 1a). The time of the reaction cycle was chosen to be 190 seconds, as was measured in our experiment. The initial curvature was chosen to be $\bar{\kappa}_r = 1 \text{ cm}^{-1}$. After 2 seconds, the curvature started to change linearly, and after 60 seconds it stabilized on value of $\bar{\kappa}_o = -0.8 \text{ cm}^{-1}$ for 10 seconds. Then, it increased linearly for 120 seconds until it returns to its initial value.

Extracting the values of D and θ

For the experiments of the passive PVS ribbons, a theoretical curve based on Eq. 9 was fitted to each velocity profile, with two free parameters, D the linear drag coefficient and θ the contact angle of the meniscus. The values of the extracted parameters are summarized in Table S1.

For the experiment of the BZ gel ribbon, a theoretical curve based on Eq. 9 was fitted to each attraction velocity profile, with three free parameters, D the linear drag coefficient, θ the contact angle of the meniscus and $l_c$ the capillary length. The values of the extracted parameters are summarized in Table S2.

The velocity profile dependence on the **length** (L) of the ribbon

Our theoretical model predicts that longer ribbons experience larger attraction force, and hence move faster towards the wall. In a series of measurements that we performed with a set of elastic ribbons in different lengths (0.3 – 1 cm) and identical width, thickness, and intrinsic curvature, we saw that the absolute value of the maximal velocity Increases with the length (Fig. S2a). Furthermore, compression between the values of $|v_{max}|$ thah we measured and the values calculated using Eq. 6 (Fig. S2b) reveals that our model captures the measured trend.

The velocity profile dependence on the **Young's modulus** (Y) of the ribbon our model predicts that stiffer ribbons, meaning larger Young's modulus Y experience grater attractive force and hence gain higher velocity during their motion. This relation is linear in the limit of soft ribbons ($\bar{O} \gg 1, \kappa^* \cong \kappa_s$) as can be seen from Eq. 2 and 6. We performed the same experiment with two different kinds of PVS ribbons with different Young's moduli, and the results are shown in Fig. S3. indeed, the green ribbon ($Y = 7 \cdot 10^6 \text{ dyn/cm}^2$), which is 3.5 times stiffer than the pink one ($Y = 2 \cdot 10^6 \text{ dyn/cm}^2$), moves faster and gains maximal velocity which is about 4 time higher than the maximal velocity of the pink ribbon. Theoretical computations predict that under the conditions of the experiments the ratio between the two maximal velocities is only 2.5, and the absolute values of the velocities are smaller than those we recorded in the lab (Fig. S3b).

**Fig. S1.**

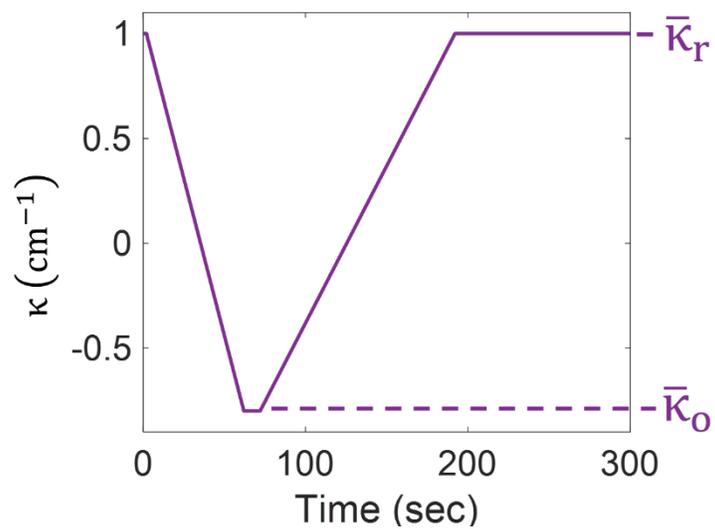

Figure S1: The intrinsic curvature of the ribbon in the modeling as a function of time. the curvature changes linearly from $\bar{\kappa}_r = 1 \text{ cm}^{-1}$ to $\bar{\kappa}_o = -0.8 \text{ cm}^{-1}$ and vice versa.

**Fig. S2.**

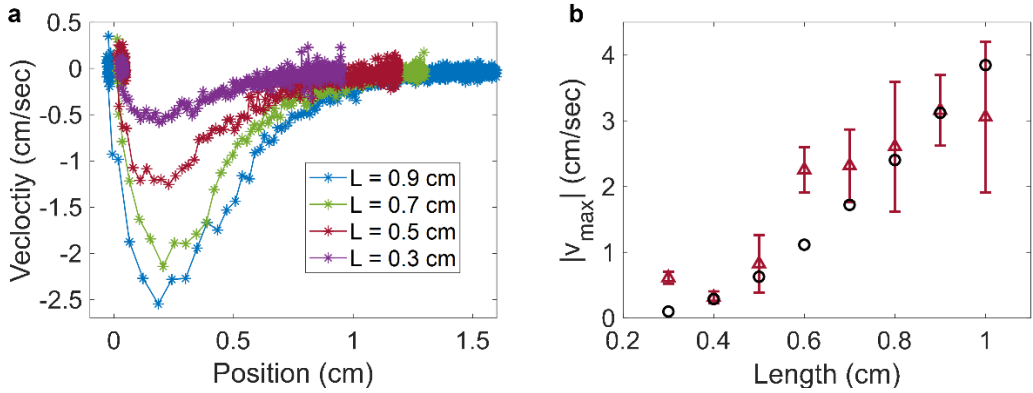

Figure S2: The effect of the ribbon's length on its motion. a) Velocity profiles of 4 different ribbons (h = 250 μm, w = 0.3 cm, $\bar{\kappa}$ = 1.27 cm$^{-1}$) floating on water (ρ = 1 g/cm$^3$, σ = 70 dyn/cm). The lengths are 0.9 cm (blue) 0.7 cm (green) and 0.5 cm (red) and 0.3 cm (purple). The longer the ribbon is, the faster it moves towards the wall. b) The absolute values of the maximal velocity of the measured ribbons as function of their lengths (red triangles) and theoretical values as computed using Eq. 6 (black circles) with the relevant experimental parameters, θ = 28° and D = 55 (dyn · sec)/cm$^2$.

**Fig. S3.**

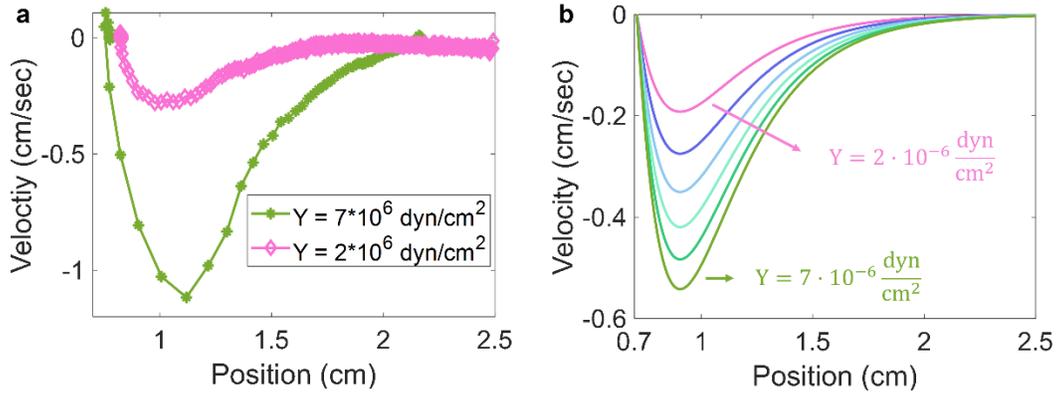

Figure S3: The effect of Youngs modulus on the motion of the ribbon. a) Velocity profiles of 2 different ribbons (L = 1 cm, h = 240 μm, w = 0.2 cm, $\bar{\kappa}$ = 0.48 cm$^{-1}$) floating on water ($\rho$ = 1 g/cm$^3$, $\sigma$ = 70 dyn/cm). The Youngs moduli are $7 \cdot 10^6$ dyn/cm$^2$ (green) and $2 \cdot 10^6$ dyn/cm$^2$ (pink). Stiffer ribbons move faster towards the wall, as was predicted by our model. b) Theoretical velocity curves of ribbons with different Young's moduli (Y = $2 - 7 \cdot 10^6$ dyn/cm$^2$) (Eq. 6) with the relevant experimental parameters and the average values of the contact angle and drag coefficients: $\theta$ = 28° and D = 55 (dyn · sec)/cm$^2$.

**Table S1.**

Table S1: Fitting parameters (D and θ) for different velocity profiles of PVS ribbon floating on water

| $\bar{\kappa}$ (cm$^{-1}$) | h (μm) | D ($\frac{dyn \cdot sec}{cm^2}$) | θ (°) |
|---|---|---|---|
| 1.05 | 290 | 44 ± 1 | 33 ± 1 |
| 1.05 | 290 | 48.9 ± 0.5 | 30.9 + 0.2 |
| 0.74 | 290 | 46.2 ± 0.6 | 29.3 ± 0.3 |
| 0.74 | 290 | 33.9 ± 0.7 | 33.7 ± 0.4 |
| 0.67 | 290 | 60.2 ± 0.7 | 26.0 ± 0.1 |
| 0.67 | 290 | 65 ± 1 | 26.1 ± 0.4 |
| 0.48 | 380 | 86 ± 4 | 20.6 ± 0.8 |
| 0.48 | 380 | 80.49 ± 3 | 21.8 ± 0.6 |
| 0.4 | 380 | 51.0 ± 0.9 | 29.3 ± 0.4 |
| 0.4 | 380 | 36 ± 2 | 29.0 ± 0.8 |

**Table S2.**

Table S2: Fitting parameters (**D**, **θ** and **l_c**) for different velocity profiles of a BZ gel ribbon floating between BZ solution and Perfluorodecalin.

| $\bar{\kappa}$ (cm$^{-1}$) | h (μm) | D ($\frac{\text{dyn} \cdot \text{sec}}{\text{cm}^2}$) | θ (°) | l$_c$ (cm) |
|---|---|---|---|---|
| 1 | 550 | 310 ± 10 | 31.3 ± 0.6 | 0.29 ± 0.01 |
| 1 | 550 | 410 ± 10 | 27.1 ± 0.9 | 0.25 ± 0.01 |
| 1 | 550 | 605 ± 7 | 21.2 ± 0.5 | 0.224 ± 0.004 |
| 1 | 550 | 455 ± 7 | 25.5 ± 0.3 | 0.255 ± 0.005 |
| 1 | 550 | 510 ± 10 | 24.8 ± 0.8 | 0.222 ± 0.008 |
| 1 | 550 | 340 ± 20 | 30.1 ± 0.6 | 0.28 ± 0.01 |
| 1 | 550 | 277 ± 8 | 31.1 ± 0.2 | 0.301 ± 0.006 |

**Movie S1.**
A BZ gel strip lying on its side and changing its color and curvature periodically due to the BZ reaction that occurs within it. The curvature changes its sign during each reaction cycle. Video played 80 times faster than real time. Gel length: 20 mm. (MOV 15.4 mb)

**Movie S2.**
A BZ gel strip surfing on the interface between Perfluorodecalin and BZ solution. Top view. The gel changes its color and curvature periodically due to the BZ reaction that occurs within it and moves from the walls of the container to its center and vice versa. Video played 100 times faster than real time. Gel length: 43 mm, container size: 2.05x3 cm. (MOV 29.1 mb)

**Movie S3.**
A BZ gel strip surfing on the interface between BZ solution and air. Side view. The gel changes its color and curvature periodically due to the BZ reaction that occurs within it and moves from the center of the container to its wall and vice versa. Video played 32 times faster than real time. Gel length: 50 mm. (MOV 26 mb)

**Movie S4.**
A BZ gel strip surfing on the interface between Perfluorodecalin and BZ solution. Top view. The gel changes its color and curvature periodically due to the BZ reaction that occurs within it and moves from the walls of the container to its center and vice versa. Video played 100 times faster than real time. Gel length: 33 mm, container size: 2.1x2.1 cm. (MOV 25 mb)

**Movie S5.**
A PVS strip surfing on the interface between water and air. Top view. The strip first turns towards the container wall and then accelerates towards it. Strip length: 70 mm. Strip curvature: $1.05$ cm$^{-1}$. (MOV 12.9 mb)

**Movie S6.**
Numerical integration of the equations of motion for a PVS strip floating on water. The strip first turns towards the container wall and then accelerates towards it. Strip length: 70 mm. Strip curvature: $1.05$ cm$^{-1}$. (MOV 1.14 mb)

**Movie S7.**
Numerical integration of the equations of motion for a BZ strip floating between Perfluorodecalin and BZ solution. The modeled strip changes its intrinsic curvature in the same manner the real gel does, over one cycle. Strip length: 43 mm. Strip curvature: $0.4$ cm$^{-1}$. (MOV 5.06 mb)